\documentclass{jfm}
\usepackage{natbib,amsmath,amssymb,graphicx}
\usepackage[usenames,dvipsnames,svgnames,table]{xcolor}
\usepackage{xcolor}
\usepackage{subcaption}
\usepackage{tikz}
\usepackage{soul}
\usepackage{pgfplots}
\ifCUPmtlplainloaded \else
\checkfont{eurm10}

\ifCUPmtlplainloaded \else
\checkfont{msam10}
\iffontfound
\IfFileExists{amssymb.sty}
{\typeout{^^JFound AMS Symbol fonts on the system, using the
		'amssymb' package.^^J}%
	\usepackage{amssymb}%
	  \let\leq=\leqslant
	  
}{}
\fi
\fi

\ifCUPmtlplainloaded \else
\IfFileExists{amsbsy.sty}
{\typeout{^^JFound the 'amsbsy' package on the system, using it.^^J}%
	\usepackage{amsbsy}}
{\providecommand\boldsymbol[1]{\mbox{\boldmath $##1$}}}
\fi
\usepackage{overpic}
\usepackage{amsmath,mathrsfs}
\usepackage{psfrag}
\usepackage{amsmath,amssymb,amsfonts,bbm}
\usepackage[bbgreekl]{mathbbol}
\usepackage{mathrsfs}
\usepackage{float}

\renewcommand{\vec}[1]{\ensuremath{\mbox{\boldmath$#1$}}}
\newcommand{\ma}[1]{\ensuremath{\mathbb{#1}}}

\newcommand\nn{\nonumber}
\newcommand{\vp}{\dot{\vec x}}

\newcommand{\uc}{u_{c}}
\newcommand{\tauc}{\tau_{c}}

\begin{document}
	
	\title{Hydrodynamic force on a small squirmer moving with a time-dependent velocity at small Reynolds numbers }
	\shorttitle{Hydrodynamic force on a small squirmer}
	\author[T. Redaelli, F. Candelier, R. Mehaddi, C. Eloy and  B. Mehlig]
	{T. Redaelli$^1$, F. Candelier$^2$, R. Mehaddi$^3$,  C. Eloy$^1$ and B. Mehlig$^4$}
	\affiliation{$^1$ Aix Marseille Univ, CNRS, Centrale Marseille, IRPHE, Marseille, France \\
		$^2$Aix-Marseille Univ.,  CNRS, IUSTI,
		Marseille, France\\
		$^3$  Université de Lorraine, CNRS, LEMTA, F-54000 Nancy, France\\
		$^4$ Department of Physics, Gothenburg University,
		SE-41296 Gothenburg, Sweden\\}
	\date{}
	\maketitle

\begin{abstract}
We calculate the hydrodynamic force on a small spherical, unsteady squirmer moving with a time-dependent velocity in a fluid at rest, taking   into account convective and unsteady fluid inertia effects in perturbation theory. Our results generalise those of Lovalenti \& Brady (1993) from passive to active spherical particles. We find that   convective inertia changes the history-contribution to the hydrodynamic force, as  it does for passive particles. We determine how the hydrodynamic force depends on the swimming gait of the unsteady squirmer. Since swimming breaks the spherical symmetry of the problem, the force is not completely determined by the outer solution of the asymptotic matching problem, as it is for passive spheres. There are additional contributions due to the inhomogeneous solution of the inner problem.  We also compute the disturbance flow, illustrating convective and unsteady effects  when the particle experiences a  sudden start  followed by a sudden stop.
 \end{abstract}

\section{Introduction}
A small motile organism swimming in a marine environment experiences a hydrodynamic force. How does this force depend on the mechanism of propulsion, upon the swimming gait of the organism? How does  the shape of the swimmer affect the hydrodynamic force, and how does it depend on the centre-of-mass velocity, and upon the angular velocity of the swimmer? Closely related questions
concern the disturbance produced by the organism as it moves through the fluid. How does the disturbance decay away from the swimmer, how does it change as a function of time for a time-dependent swimming gait, and how do sudden accelerations affect the disturbance?
 These are important questions, because the disturbance flow determines hydrodynamic interactions between organisms, and it is known that 
 motile microorgansisms  employ hydrodynamic signals to localise prey, or to escape their predators, or to find mates  \citep{guasto2012fluid}.

Answering these questions in  general is obviously a very challenging task. To simplify the problem, the dynamics of motile microorganisms in water is often described using the creeping-flow  approximation -- the Stokes approximation -- which neglects possible effects of  unsteady and convective fluid inertia \citep{Lighthill1952,Blake1971, fauci2006, yates1986, happelbrenner1981,pedleysquirmer,lauga2009,Visser2011}.  This approach works very well  for the majority of microorganisms. But for larger ones,  in particular for organisms that swim with  highly unsteady time-dependent gaits,  fluid inertia cannot be neglected.  
The flow field of the organism {\em Chlamydomonas reinhardtii}, for instance, cannot be described by the Stokes
approximation \citep{wei2021measurements}.
As a second example, consider  {\em Mesodinium rubrum} which swims with short jumps interrupted by longer rest periods \citep{fenchel2006motile,Jiang2011Robidium}. When the organism stops suddenly, fluid inertia cannot be neglected:
the Stokes approximation predicts that the surrounding fluid arrests instantly. 
In reality, the disturbance flow around the organism takes some time to vanish. For a passive sphere, this inertia effect induced by the unsteadiness of the disturbance flow gives rise to a memory term in the expression for the hydrodynamic force that depends on its past accelerations, the BBO history force  \citep{boussinesq1885resistance,basset1888treatise,oseen1927neuere}. 
  \citet{wang2012unsteady}  used the unsteady Stokes equation to model how the velocity
of a  spherical squirmer decays after a sudden jump to escape a predator. They determined how the history force affects the velocity decay, and found good agreement with measurements performed by  \citet{Jiang2011} on copepods, despite the fact that possible effects of convective
fluid inertia were not considered. 
 \cite{spelman2017arbitrary} calculated the hydrodynamic force and the disturbance flow produced by a deforming swimmer, accounting
 for unsteady fluid inertia.
 
At least for passive particles, it is known that convective fluid inertia,  induced by a non-zero slip velocity between the organism and the surrounding fluid, 
 can change the long-time behaviour of the history force. For times larger than 
the Oseen time,
 the kernel of the history force decays more rapidly than the BBO kernel, namely as $t^{-2}$ for a sudden start \citep{Sano1980history,lovalenti1993hydrodynamic} instead of the $t^{-1/2}$-decay of the BBO history force. 
\citet{Sano1980history} and \citet{lovalenti1993hydrodynamic}  derived their results for weak convective fluid inertia, 
  for small but non-zero particle Reynolds numbers Re$_{\rm{p}}$. In the experiments carried out by \citet{Jiang2011}, Re$_{\rm{p}}$ was quite high, however. The 
 small-Re$_{\rm{p}}$ theory is expected to fail in this case.

It is not known how the results of  \citet{Sano1980history} and \citet{lovalenti1993hydrodynamic}
generalise to motile organisms, yet fluid-inertia forces  are thought to play a significant role for swimmers in many situations. 
\cite{wang2012inertial}, for instance, determined convective-inertia corrections for a steady, spherical squirmer, modelling the swimming gait as 
a steady surface velocity \citep[see also][]{ khair2014expansions,Chisholm2016}. However, the
the slip velocity of  motile organisms is usually time dependent, and the quasi-steady approximation may fail if 
the slip velocity varies much faster than the disturbance flow.

In order to understand how convective inertia modifies the unsteady dynamics of a small motile organism, we calculated the 
hydrodynamic force on a small spherical squirmer in an unsteady, spatially homogeneous flow, thus generalising the results of  \citet{lovalenti1993hydrodynamic} to a small motile organism. 
 While \citet{lovalenti1993hydrodynamic}  used the reciprocal theorem to obtain their results, we employed  asymptotic matching of perturbation expansions \citep{Hinch1995} 
in the parameter $\varepsilon = \sqrt{{\rm Re}_{\rm{p}} {\rm Sl}}$, where ${\rm Sl}$ is the Strouhal number. 
We note that \citet{sennitskii1990self} computed the disturbance velocity for a swimmer, far from its surface, using asymptotic matching of
perturbation expansions (in a different parameter),  but not the hydrodynamic force. Our solution  is related to that of
 \citet{lovalenti1993hydrodynamic}. In particular, we find that the hydrodynamic force on the squirmer involves memory forces, 
 and that their kernels are closely related to those of  \citet{lovalenti1993hydrodynamic}. For all examples we checked, the kernels were in fact numerically identical.
 
 We identified two major differences to the passive case. First, for the squirmer, the memory forces involve a source term that contains an active part. Second, the inhomogeneous part of the solution to the inner problem of order $\varepsilon$ contributes to the hydrodynamic force on the squirmer. For a passive spherical particle, by contrast, spherical symmetry ensures that such contributions vanish. We discuss the significance of the inhomogeneous contribution for the squirmer.  Last but not least, asymptotic matching allows us to determine the disturbance flow. We show how the disturbance develops  for a sudden start and for a sudden stop, and discuss the implications of our findings for the biology of small motile organisms in a marine environment.

\section{Formulation of the problem}
	\label{sec:model}
\begin{figure}
	\centering
	\begin{overpic}[width=2.8cm,clip]{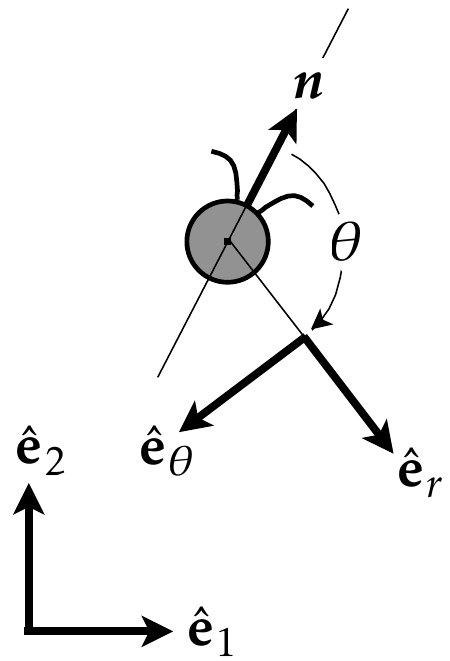}
	\end{overpic}
	\caption{\label{fig:schematic} Schematic illustration of the squirmer model described in Section~\ref{sec:model}.  The squirmer swims head first, along the direction $\vec n$
	in the $\hat{\bf e}_1$-$\hat{\bf e}_2$-plane.
	Points on the surface in this plane are parameterised by the angle $\theta$. Shown are the lab-coordinate system $\hat{\bf e}_i$ and the coordinate system $\hat{\bf e}_r$
	and $\hat{\bf e}_\theta$ in the body frame.}
\end{figure}

The spherical squirmer is  an idealised model for a microswimmer, introduced by \cite{Lighthill1952} and \cite{Blake1971}. It is  widely used to investigate the dynamics of motile microorganisms, their interactions, and collective behaviours \citep{pedleysquirmer,lauga2009,Visser2011,Pak2014}.  In this model, the swimming gait is represented by an 
axisymmetric surface-velocity field in the frame translating with the squirmer
\begin{align}
\vec v (\theta,t) = \sum_l A_l(t) P_l(\cos\theta) \hat{\bf e}_r+ B_l(t) {V_l}(\cos\theta)  \hat{\bf e}_\theta\,.
\end{align}
Here $\theta$ parameterises points on the surface, and $\hat{\bf e}_r$ and $\hat{\bf e}_\theta$
are the corresponding basis vectors in the body frame (Figure \ref{fig:schematic}). Using the notation of \cite{Lighthill1952},
$P_l$ is the Legendre polynomial of order $l$, and $V_l$ is related to the first associated Legendre polynomial of order $l$,
namely $V_l = -2 P_l^1/[l (l+1)]$. In general, the
coefficients  $A_l(t)$ and $B_l(t)$ are allowed to depend on time, but  the simplest version of the model is the steady, spherical squirmer with  a
time-independent, tangential surface-velocity field \citep{pedleysquirmer}
\begin{align}
	\vec v(\theta)= \big(B_1\sin \theta + B_2\sin \theta \cos \theta \big) \hat{\bf e}_\theta\:.
	\label{eq:ActiveSpeed}
\end{align}
When inertial effects are negligible, one solves the steady Stokes equation with the boundary condition
 (\ref{eq:ActiveSpeed}) to determine the hydrodynamic force acting on such a spherical squirmer. The result is
\begin{equation}
\vec f'^{(0)} = -6 \pi \mu a \Big( \dot{\vec x} -\tfrac{2}{3}B_1 \vec n\Big)\:,
\label{Stokes_force}
\end{equation} 
where $\mu$ is the dynamic viscosity of the fluid, $\vec n$ is the unit vector along the symmetry axis of the surface-velocity field
 (Figure \ref{fig:schematic}), $a$ is the radius of the spherical squirmer, and $\vec x$ the position vector of its centre of mass in the laboratory frame. 
 The steady swimming velocity in the Stokes approximation is obtained by setting the hydrodynamic force to zero. This gives
 $\vp= (2/3) B_1 \vec n$.  

We consider an {\em  unsteady} squirmer,  with time-dependent coefficients $B_1(t)$ and $B_2(t)$ in a quiescent fluid.
The surface-velocity field produces a time-dependent disturbance flow that causes
the organism to move with a  time-dependent  centre-of-mass velocity $\vp(t)$.
In order to determine the hydrodynamic force on the squirmer, one has to solve 
	the continuity and momentum equations for the disturbance flow $\vec{w}$ and the disturbance pressure $p$. In non-dimensional variables, these equations take the standard form 
	\begin{subequations}
	\label{eq:eom}
	\begin{align}
	\boldsymbol{\nabla} \cdot \vec{w} &= 0\:,
	\label{div_w}\\
	{\rm Re}_{\rm{p}}\,{\rm Sl} \frac{\partial \vec{w}}{\partial t} \Big|_{\vec{r}} -{\rm Re}_{\rm{p}} \vp \cdot \boldsymbol{\nabla} \vec{w}+{\rm Re}_{\rm{p}} \: \vec{w} \cdot \boldsymbol{\nabla} \vec{w} &= - \boldsymbol{\nabla} p + \boldsymbol{\triangle} \vec{w} \:.
	\label{eq_mvt}
	\end{align}
	When the angular velocity  of the swimmer is negligible, the boundary conditions for Eq.~(\ref{eq_mvt}) read:
	\begin{equation}
	\vec{w} = \vp + \vec{v}\,\,\mbox{for}\,\,|\vec r|=1\,,  \quad \mbox{and} \quad \vec{w} \to \vec{0}\,\,\mbox{as}\,\, |\vec r|\to \infty\,.
	\label{eq_Bc}
	\end{equation}
	\end{subequations}
	All vectors in (\ref{eq:eom}) are expressed in the laboratory frame, but using a system of moving coordinates that translates with the  squirmer. In particular, Eq. (\ref{eq_mvt}) is obtained  using the following relation that links the partial time derivative at a fixed point $\vec{x}$ in the laboratory frame to the partial time derivative at a fixed point $\vec{r}$ in the moving coordinate system: $\partial/\partial t\big|_{\vec x} = \partial/\partial  t\big|_{\vec r}-\dot{\vec x} \cdot \vec \nabla$.
	Here $\dot{\vec x}$ is the centre-of-mass velocity of the squirmer.
	Furthermore, we non-dimensionalised lengths by dividing with the radius $a$ of the spherical squirmer, 
	 velocities with a typical velocity $\uc$, and time with a typical time scale  $\tauc$. 
The resulting non-dimensional parameters are the particle Reynolds number and the Strouhal number:
	\begin{align}
	{\rm Re}_{\rm{p}} = \frac{a \uc}{\nu} \quad \mbox{and} \quad {\rm Sl} = \frac{a}{\uc \tauc}\:. 
	\end{align}
	The particle Reynolds number determines the importance of convective terms in Eq.~(\ref{eq_mvt}), and ${\rm Re}_{\rm{p}} {\rm Sl}=a^2/(\nu \tauc)$ is 
	the ratio of two time scales, the viscous time $\tau_\nu=a^2/\nu$, and $\tauc$. Equations~(\ref{div_w},\ref{eq_mvt})  are precisely  those used by \citet{lovalenti1993hydrodynamic}, 
	to study the effect of convective inertia on the hydrodynamic force on a passive sphere in a spatially uniform flow, their Eqs.~(2.3a, 2.3b). 
	Only the boundary conditions (\ref{eq_Bc}) differ, by the surface-velocity field
	$\vec v$. This additional term makes the difference between an active and a passive particle. 
	In the following we describe how this term modifies the hydrodynamic force.
	
\section{Matched asymptotic expansions}
	\label{sec:flow_field}
We solve Eqs.~(\ref{eq:eom}) under the assumption that
		\begin{align}
		\label{eq:ndp}
	\rm{Re}_{\rm{p}} \ll1 \quad \mbox{and} \quad \rm{Re}_{{p}}\rm{ Sl} \ll 1\:.
	\end{align}
	In this limit, the disturbance near the spherical  squirmer is well approximated by the quasi-steady Stokes solution which decays as $|\vec r|^{-1}$ for large values of $|\vec r|$.  As a consequence, the unsteady term $\rm{Re}_{\rm{p}}\:\rm{Sl}\: \partial_t \vec{w}$ in Eq. (\ref{eq_mvt})  becomes of the same order of magnitude as the viscous term $\boldsymbol{\triangle} \vec{w}$,
	  at a distance $|\vec r| \sim \ell_p\equiv a /\sqrt{{\rm Re}_{\rm{p}}\,{\rm Sl}}$ called the 'penetration length'. The convective term ${\rm Re}_{\rm{p}} \:\dot{ \vec{x}} \cdot \boldsymbol{\nabla} \vec{w}$ in Eq. (\ref{eq_mvt}) is of the same order of magnitude as the viscous term at the Oseen length, $|\vec r| \sim \ell_{O} \equiv a /{\rm Re}_{\rm{p}}$.
	  
	  The problem has two asymptotic limits. In the limit ${\rm Sl}\to 0$, the problem becomes steady, leading to an Oseen correction 
	  to the hydrodynamic force \citep{wang2012inertial,khair2014expansions,Chisholm2016}. Conversely, when ${\rm Re}_{\rm{p}}\to 0$ and ${\rm Sl}\to \infty$, 
	  keeping $\varepsilon^2\equiv{\rm Re}_{\rm{p}}\,{\rm Sl}$ constant, one obtains an unsteady Stokes problem. Its solution yields the BBO history force obtained by \cite{wang2012unsteady}. 
	
	Since  inertial corrections are singular perturbations,   we use asymptotic matching of perturbation expansions  \citep{Hinch1995} in the parameter $\varepsilon{=\sqrt{{\rm Re}_{\rm{p}}\rm{Sl}}}$ to compute the inertial corrections to the hydrodynamic force on the squirmer. 
	  It is natural
	to take the independent non-dimensional parameters as:
	 \begin{equation}
	 \label{eq:new_params}
	\varepsilon = \sqrt{{\rm Re}_{\rm{p}}\rm{Sl}} \quad \mbox{and}\quad   {\ell_p}/{\ell_{O}} =\sqrt{ {\rm Re}_{\rm{p}}/\rm{Sl}}\,.
	\end{equation}
	We require $\varepsilon$ to be small -- in keeping with (\ref{eq:ndp}) -- and  ${\ell_p}/{\ell_{O}}$  to remain finite as $\varepsilon$ becomes small.
	The length-scale ratio  ${\ell_p}/{\ell_{O}}$  characterises the competition between convective and unsteady inertia upon the disturbance flow.
        At first sight, it might appear that one cannot treat large unsteadiness in this fashion,  but we show below that the hydrodynamic force obtained in this way is uniformly
         valid. 
	In terms of the parameters (\ref{eq:new_params}), the equations of motion (\ref{eq:eom}) take the form
	\begin{subequations}
		\label{eq_mvt2}
	\begin{align}
	\boldsymbol{\nabla} \cdot {\vec{w}}&= \vec{0}\:, \\
	\label{eq_mvt32}
	\varepsilon^2 \frac{\partial \vec{w}}{\partial t} \Big|_{\vec{r}} - \varepsilon \left(\frac{\ell_p}{\ell_{O}} \right)\vp \cdot \boldsymbol{\nabla} \vec{w}+\varepsilon \left(\frac{\ell_p}{\ell_{O}} \right) \: \vec{w} \cdot \boldsymbol{\nabla} \vec{w} &= - \boldsymbol{\nabla} p + \boldsymbol{\triangle} \vec{w} \,.
\end{align}
\end{subequations}
To find the disturbance flow $\vec w$, 
configuration space is divided in two different regions, the inner region, $|\vec r| \sim 1$, and the outer region  $|\vec  r| \gg 1$.  In these two regions, disturbance flow and  pressure  are written in the form of two different  series expansions in $\varepsilon$. The inner and outer expansions are  matched at $|\vec r| \sim 1/\varepsilon$. 
This yields the necessary boundary conditions
	for the inner problem which can then be solved to determine the hydrodynamic force \citep{Hinch1995}.

Before going further,  let us briefly comment on the different terms on the left-hand side of Eq. (\ref{eq_mvt32}).  
Near the  squirmer, all terms are small when $\varepsilon$ is small. But while the unsteady term scales as $\varepsilon^2$,  the convective terms are proportional to $\varepsilon$.
These terms cause the key difficulty. When the perturbation is of order $\varepsilon^n$ with $n>1$, the hydrodynamic force is given
by the outer solution alone \citep{redaelli2021,legendre1998lift}. In the present case, the perturbation is larger -- of order $\varepsilon$ --
and therefore the   method of \citet{redaelli2021} cannot be applied here. As a consequence, we need to consider the details of the inner solution. We discuss the outer solution first, however, because it yields
the boundary conditions needed for solving the inner problem.
 
\subsection{Outer solution}
Far from the  squirmer, the boundary conditions on the  surface of the organism can be replaced by a  Dirac delta-function with amplitude $\vec f^{(0)}$ \citep{Childress64}:
\begin{subequations}
\label{eq:eom52}
\begin{align}
	\boldsymbol{\nabla} \cdot {\vec{w}}_{\mbox{\scriptsize out}}&= \vec{0}\:, 
	\label{div_w2}\\
	\varepsilon^2 \frac{\partial \vec{w}_{\mbox{\scriptsize out}}}{ \partial t} \Big|_{\vec{r}} -\varepsilon \left(\frac{\ell_p}{\ell_{O}}\right)\vp \cdot \boldsymbol{\nabla} {\vec{w}}_{\mbox{\scriptsize out}}  & = - \boldsymbol{\nabla} {p}_{\mbox{\scriptsize out}}  + \boldsymbol{\triangle}{\vec{w}}_{\mbox{\scriptsize out}}  +  \vec{f}^{(0)} \delta(\vec{r}) \:.
	\label{eq:outer_a}
	\end{align}
In the matching region, the quadratic term $\varepsilon (\ell_p/\ell_{O})  \: \vec{w} \cdot \boldsymbol{\nabla} \vec{w} $ in (\ref{eq_mvt2}) is negligible compared to the other terms in the equation, because its magnitude scales as $\varepsilon (\ell_p/\ell_{O})|\vec r|^{-3}\sim \varepsilon^4$. 
The amplitude of the source term  is the opposite of the Stokes force (\ref{Stokes_force}),  
\begin{align}
\vec f^{(0)}= -{6\pi}({\tfrac{2}{3} B_1 \vec n} -  \vp)\,.
\label{eq:f0_def}
\end{align}
	\end{subequations}
The first term on the r.h.s.
is an active part, related to the surface-velocity field $\vec v$.

The solution of the outer problem (\ref{eq:eom52}) is derived in the Appendix. It reads
 \begin{subequations}
 \label{eq:outer_solution}
\begin{align}\label{W_outer}
	{\vec{w}}_{\mbox{\scriptsize out}} &
	 = \underbrace{\frac{1}{8\pi}\left(\frac{\ma{1}}{r}  + \frac{\vec{r}\otimes \vec{r}}{r^3}\right) \cdot  \vec{f}^{(0)}}_{{\equiv\vec{\mathcal{T}}_\mathrm{reg}^{(0)}(\vec{r},\:t)}}
	 \underbrace{-{\rm Re}_{\rm{p}} \vp\cdot \boldsymbol{\nabla}\left[\frac{3 r}{32\pi} \left( \ma{1} - \frac{1}{3} \frac{\vec{r} \otimes \vec{r}}{r^2}\right) \cdot \vec{f}^{(0)} \right]}_{{\equiv\varepsilon \:\vec{\mathcal{T}}_\mathrm{reg}^{(1)}(\vec{r},\:t)}}\\
	&  
	\underbrace{-  \varepsilon \int_{0}^{t}\frac{\mbox{d}\tau}{6\pi}\, {\ma{K}^{(1)}(t,\tau)}\cdot \tfrac{{\rm d}}{{ \rm d} \tau} \vec{f}^{(0)}(\tau)  - {\rm Re}_{\rm{p}} \: \int_{0}^{t}\frac{\mbox{d}\tau}{6\pi} \,\ma{ K}^{(2)}(t,\tau) \cdot \vec{f}^{(0)}(\tau)}_{{\equiv\varepsilon  \vec {\mathcal U}(t)}} \,.
		\nn
		\end{align}
The integral kernels $\ma{K}_1$ and $\ma{K}_2$ in Eq.~(\ref{W_outer}) have elements:
\begin{align}
	\label{eq:K1}
	[\ma{K}&^{(1)}(t,\:\tau)]_{ij} 
	= - \tfrac{3}{8} \Big[ \big(\!-2\!+\!A(t,\:\tau)^{-2}\big) \frac{\mbox{erf}(A(t,\:\tau))}{A(t,\:\tau)}- 2\frac{\exp(-A(t,\:\tau)^2)}{A(t,\:\tau)^2\sqrt{\pi}} 
	\Big] \frac{\delta_{ij}}{\sqrt{t-\tau}}  \\\nn
	&
	- \tfrac{3}{8} \Big[ \big(1\!-\!\tfrac{3}{2}A(t,\:\tau)^{-2}\big) \frac{\mbox{erf}(A(t,\:\tau))}{A(t,\:\tau)}+ 3\frac{\exp(-A(t,\:\tau)^2)}{A(t,\:\tau)^2\sqrt{\pi}} \Big] \frac{(\delta_{ij} - q_i(t,\tau) q_j(t,\tau))}{\sqrt{t-\tau}}\,,\\
\label{eq:K2}
	[\ma{K}&^{(2)}(t,\:\tau)]_{ij}   = -\tfrac{3 }{8} \frac{1}{2A(t,\:\tau)}\Big[ \big(1-\tfrac{3}{2}A(t,\:\tau)^{-2}\big) \frac{\mbox{erf}(A(t,\:\tau))}{A(t,\:\tau)}+ 3\frac{\exp(-A(t,\:\tau)^2)}{A(t,\:\tau)^2\sqrt{\pi}} \Big] \\\nn
	& \times \frac{q_i(t,\tau) {\dot x}_j(\tau) - 3q_i(t,\tau) q_j(t,\tau) {\sum_{k=1}^3}q_k(t,\tau) {\dot x}_{k}(\tau) }{t-\tau}\\\nn
	&- \tfrac{3 }{8} \Big( \frac{1}{2A(t,\tau)}\Big)\Big[ 3\frac{\mbox{erf}(A(t,\tau))}{A(t,\tau)^3}-\big(4+6 A(t,\tau)^{-2}\big)\frac{\exp(-A(t,\tau)^2)}{\sqrt{\pi}} \Big] \\\nn
	&
	\times \frac{  [\delta_{ij} - q_i(t,\tau) q_j(t,\tau)]{\sum_{k=1}^3}q_k(t,\tau){\dot x}_{k}(\tau)}{t-\tau}\,.
\end{align}
	Here $A(t,\:\tau)$ is the norm of the  {\em pseudo-displacement} vector  $\vec{a}(t,\:\tau)$ introduced by \cite{lovalenti1993hydrodynamic}  in Eq.~(6.7d) of their paper: 
	\begin{equation}
	\vec{a} (t,\:\tau) \equiv \left(\frac{\ell_p}{\ell_{O}}\right)  \frac{1}{2 \sqrt{t-\tau}} \int_\tau^t \!\!\mbox{d}t' \,\dot{\vec x}(t')\:,
	\label{eq:a}
	\end{equation}
	\end{subequations}
and  $q_i$ are the components of the (unit) direction vector $\hat{\vec{q}}= \vec{a}(t,\:\tau)/A(t,\:\tau)$.  
We note that the outer solution derived here differs from that obtained by \cite{shu2001generalized}; 
 they considered a Dirac delta-function force $\delta(t)$, instead of a continuously time dependent force. 

Let us briefly discuss the different terms in the outer solution (\ref{W_outer}). 
Following \citet{Meibohm2016}, we denote
the first term on the right-hand side of Eq.~(\ref{W_outer}) by  $\vec{\mathcal{T}}_\mathrm{reg}^{(0)}(\vec{r},\:t) $. This is the well-known Stokeslet describing
the far-field disturbance flow produced by an active particle in the Stokes limit.
The  second term on the right-hand side of Eq.~(\ref{W_outer})  is denoted by $\varepsilon \:\vec{\mathcal{T}}_\mathrm{reg}^{(1)}(\vec{r},\:t)$.
This term represents a regular perturbation to the Stokeslet $\vec{\mathcal{T}}_\mathrm{reg}^{(0)}(\vec{r},\:t)$.
The two integral terms, finally,  combine to give the history force. It can be written in the form
$\varepsilon  \vec {\mathcal U}(t)$, where $\vec {\mathcal U}(t)$ is a spatially uniform flow at infinity. 
A passive sphere has $B_1\!=\!B_2\!=\!0$. In this case, spherical symmetry ensures that all first-order inertia corrections to the hydrodynamic force
are due to the outer flow alone. In this case, one obtains $\vec{f}'= \vec{f}'^{(0)} + 6\pi \varepsilon \vec{\mathcal{U}}(t)$ to order $\varepsilon$. This expression for the hydrodynamic force is equivalent to that obtained by \cite{lovalenti1993hydrodynamic} [their Eq.~(6.15)] with three minor differences. First,   \citet{lovalenti1993hydrodynamic} combined the terms multiplying $\vec f^{(0)}$ and the time derivative of $\vec f^{(0)}$. We kept them separate here,  because they  have different behaviours at short and long times 
(Section~\ref{sec:results}).
Second, we assumed  that $\vec f^{(0)}=\vec 0$ for $t \leq 0$. As a consequence, the integration domain is $[0,t]$ instead of $[-\infty,t]$.
Third,  our expression for the kernel looks slightly different from the kernel in Eq.~(6.15) of \cite{lovalenti1993hydrodynamic}.
However, we found that the two expressions are numerically equivalent for
all centre-of-mass motions we examined: sudden start \citep{Sano1980history}, linearly increasing velocity $\vp=\vec v_0 t/t_0$, and sinusoidally varying  centre-of-mass velocity $\vp =\vec v_0 {\rm \sin}(\omega_0 t)$, with coefficients $\vec v_0$, $t_0$, and $\omega_0$.  It is quite common that the two methods, reciprocal theorem and asymptotic matching, yield expressions for the hydrodynamic force and torque that look different but are equivalent~\citep{Meibohm2016}. 

\subsection{Inner solution}
Near the  squirmer, for $| \vec r|  = O(1)$, the disturbance flow is expanded as
\begin{equation}
{\vec{w}}_{\mbox{\scriptsize in}} = {\vec{w}}_{\mbox{\scriptsize in}}^{(0)} + \varepsilon  \: {\vec{w}}_{\mbox{\scriptsize in}}^{(1)} +  \mathcal{O}(\varepsilon^2)
	\quad \mbox{and} \quad  
	{p}_{\mbox{\scriptsize in}} = {p}_{\mbox{\scriptsize in}}^{(0)} + \varepsilon  \: {p}_{\mbox{\scriptsize in}}^{(1)} +   \mathcal{O}(\varepsilon^2) \:.
	\label{eq:flow_plotted}
\end{equation}
These inner expansions must be matched, order by order,  to the outer expansion  (\ref{W_outer}). 	
	At order $\varepsilon^0$, the inner problem to solve is
	\begin{subequations}
	\label{eq:eps0}
	\begin{align}
	&\boldsymbol{\nabla} \cdot  {\vec{w}}_{\mbox{\scriptsize in}}^{(0)}= \vec{0}\,,\quad
	- \boldsymbol{\nabla} {p}_{\mbox{\scriptsize in}}^{(0)} + \boldsymbol{\triangle}  
	{\vec{w}}_{\mbox{\scriptsize in}}^{(0)}  =  \vec{0} \:,
	\label{w0_0}\\
	&{\vec{w}}_{\mbox{\scriptsize in}}^{(0)} =   \vp (t) + \vec{v}(t)  \: \:\mbox{for}\:\: { |\vec r|=1}
	\quad \mbox{and} \quad 
	\quad {\vec{w}}_{\mbox{\scriptsize in}}^{(0)} \sim  \vec{\mathcal{T}}_\mathrm{reg}^{(0)} \: \:\mbox{for}\:\:  |\vec r| \to \infty\:.
	\label{w_0_BC}
	\end{align}
	\end{subequations}
	This is a homogeneous Stokes problem. Its solution is well known \citep{Blake1971}.
	At order $\varepsilon$, an inhomogeneous Stokes problem must be solved:
	\begin{subequations}
	\label{eq:eps1}
\begin{align}
&	\boldsymbol{\nabla} \cdot  {\vec{w}}_{\mbox{\scriptsize in}}^{(1)}=  0\,,\:\:\:\:
	- \boldsymbol{\nabla} {p}_{\mbox{\scriptsize in}}^{(1)} + \boldsymbol{\triangle}  
	{\vec{w}}_{\mbox{\scriptsize in}}^{(1)}  = - \left(\frac{\ell_p}{\ell_{O}}\right) \dot{\vec{x}}\cdot \boldsymbol{\nabla} {\vec{w}}_{\mbox{\scriptsize in}}^{(0)} +  \left(\frac{\ell_p}{\ell_{O}}\right) {\vec{w}}_{\mbox{\scriptsize in}}^{(0)}\cdot \boldsymbol{\nabla} {\vec{w}}_{\mbox{\scriptsize in}}^{(0)}\:, 
	\label{eq_order1}
	\\&
	{\vec{w}}_{\mbox{\scriptsize in}}^{(1)} =   \vec{0}    \: \:\mbox{for}\:\: { |\vec r|=1}
	\quad \mbox{and} \quad 
	\quad {\vec{w}}_{\mbox{\scriptsize in}}^{(1)} \sim  \vec{\mathcal{T}}_\mathrm{reg}^{(1)} + \vec{\mathcal{U}}(t) \:\:\mbox{for}\:\:  |\vec r|\to \infty\:. 
	\label{eq_order_1_BC}
	\end{align}
	\end{subequations}
	Since the problem is inhomogeneous, we seek its solution
	 in the form of a sum of a particular and the homogeneous solution:
	\begin{equation}
	{\vec{w}}_{\mbox{\scriptsize in}}^{(1)} = [ {\vec{w}}_{\mbox{\scriptsize p}}^{(1)} + \vec{\mathcal{U}}(t) ] + {\vec{w}}_{\mbox{\scriptsize h}}^{(1)}  \quad \mbox{and} \quad \:p_\mathrm{in}^{(1)}  = \:p_\mathrm{p}^{(1)}+\:p_\mathrm{h}^{(1)}\:.
	\end{equation}
        Consider first the particular solution.
	The velocity  ${\vec{w}}_{\mbox{\scriptsize p}}^{(1)}$ and the pressure $p_\mathrm{p}^{(1)}$ must satisfy
		\begin{equation}
	\boldsymbol{\nabla} \cdot  {\vec{w}}_{\mbox{\scriptsize p}}^{(1)}=  0\,,\quad
	- \boldsymbol{\nabla} {p}_{\mbox{\scriptsize p}}^{(1)} + \boldsymbol{\triangle}  
	{\vec{w}}_{\mbox{\scriptsize p}}^{(1)}  = - \left(\frac{\ell_p}{\ell_{O}}\right) \dot{\vec{x}}\cdot \boldsymbol{\nabla} {\vec{w}}_{\mbox{\scriptsize in}}^{(0)} +  \left(\frac{\ell_p}{\ell_{O}}\right) {\vec{w}}_{\mbox{\scriptsize in}}^{(0)}\cdot \boldsymbol{\nabla} {\vec{w}}_{\mbox{\scriptsize in}}^{(0)}\:.
	\label{eq_order1_b}
	\end{equation}
	    Since we added the uniform term $\vec {\mathcal U}(t)$ to  ${\vec{w}}_{\mbox{\scriptsize p}}^{(1)}$, the boundary condition at infinity is
	 ${\vec{w}}_{\mbox{\scriptsize p}}^{(1)} \sim \vec{\mathcal{T}}_\mathrm{reg}^{(1)}$ as $|\vec r|\to \infty$. 
	 The  inhomogeneous Eq. (\ref{eq_order1_b}) can be solved using Fourier transform \citep{candelier2023second}. 
	 The homogeneous part of the full solution satisfies:
	\begin{subequations}
	\label{eq:st_problem_2}
	\begin{align}
	&\boldsymbol{\nabla} \cdot  {\vec{w}}_{\mbox{\scriptsize h}}^{(1)}=  0\,,\quad
	- \boldsymbol{\nabla} {p}_{\mbox{\scriptsize h}}^{(1)} + \boldsymbol{\triangle}  
	{\vec{w}}_{\mbox{\scriptsize h}}^{(1)}  = \vec{0}\:, 
	\label{eq_order1_c}\\&
	{\vec{w}}_{\mbox{\scriptsize h}}^{(1)} =    - {\vec{w}}_{\mbox{\scriptsize p}}^{(1)}\Big|_{|\vec r|=1} - \vec{\mathcal{U}}(t) \:, \quad  |\vec r|=1
	\quad \mbox{and} \quad 
	\quad {\vec{w}}_{\mbox{\scriptsize in}}^{(1)} \sim  \vec{0} \:, \quad r \to \infty\:. 
	\label{boundary_p_stokes}
	\end{align}
	\end{subequations}
This Stokes problem is solved by Lamb's general solution \citep{happelbrenner1981}.
 Integrating the corresponding stress tensor over the surface of the swimmer yields the hydrodynamic force. For a passive spherical particle, the contribution of the inhomogeneous part
 of the inner solution to the hydrodynamic force must vanish due to spherical symmetry. In this case, the inertial corrections to the hydrodynamic force are determined by the outer solution alone. For the swimmer, the inhomogeneous solution contributes to the hydrodynamic force. This explains why the method of \citet{redaelli2021}  fails to determine the entire corrections to the hydrodynamic force due to convective-inertia effects. 
 The solutions of (\ref{eq:eps0}) and (\ref{eq:eps1}) obtained in this way are quite lengthy, we do not reproduce them here. 

\section{Results}
\label{sec:results}
\subsection{Hydrodynamic force}
\label{section_Hydro_force}
For the unsteady squirmer, the calculations outlined in the previous Section yield the hydrodynamic force
	\begin{align}	\label{force_re}
	\vec f' = &   - 6 \pi [\vp  -\tfrac{2}{3} B_1(t)\vec n(t) ]\\\nonumber
	&-  \varepsilon \, 6\pi \int_{0}^{t} \!\mbox{d}\tau \, \ma{K}^{(1)}(t,\tau) \cdot \frac{\mbox{d}}{\mbox{d}\tau}
	[ \vp(\tau)  -\tfrac{2}{3} B_1(\tau)\vec n(\tau)] 
  \\
	& \nonumber
	-  {\rm{Re}_{\rm{p}} } \,6\pi  \int_{0}^{t} \!{\mbox{d} \tau\:\ma{K}^{(2)}(t,\tau) } \cdot [\vp(\tau)  -{\tfrac{2}{3}} B_1(\tau)\vec n(\tau)]   \\
	& 
	-  {\rm{Re}_{\rm{p}} } \big[ \tfrac{2 \pi}{5} \dot{x}(t) B_2(t) + \tfrac{2\pi}{15} B_1(t) B_2(t) \big]\vec n(t)  \:.\nonumber
\end{align}
This is our main result, the hydrodynamic force on an unsteady, spherical squirmer in a time-dependent homogeneous flow at small but non-zero particle Reynolds number.
Equation (\ref{force_re}) generalises Eq.~(6.15) of \citet{lovalenti1993hydrodynamic} from a passive to an active sphere. As already discussed, the numerical calculations described above indicate that the kernels in Eq. (\ref{force_re}) combine to those of \cite{lovalenti1993hydrodynamic}.

Equation~(\ref{force_re}) simplifies to known solutions when unsteadiness is either very weak or very strong. To see this, consider the asymptotic
behaviours of the integrals 
\begin{equation}
{\vec I_1=-  \varepsilon \int_{0}^{t}\mbox{d}\tau \, \ma{K}^{(1)}(t,\tau) \cdot \tfrac{{{ \rm d}}}{{ \rm d} \tau}}\vec{f}^{(0)}(\tau)\,,
{\quad\mbox{and}\quad \vec I_2= - {\rm Re}_{\rm{p}} \: \int_{0}^{t} {\rm d}\tau\,\ma{ K}^{(2)}(t,\tau) \cdot \vec{f}^{(0)}(\tau)} \,.
\end{equation}
In the steady limit,   $\ell_p / \ell_{O} \to \infty$, we find
        \begin{equation}
        \vec{I}_1 \to \vec{0}\,, \quad \mbox{and} \quad \vec{I}_2  \to - \tfrac{3}{8} \mathrm{Re_{\rm{p}}}  |\dot{\vec x}|\vec{f}^{(0)}\,.
        \label{eq:to_inertia}
        \end{equation}
Only the  convective Oseen correction survives. It combines with the contribution of the inhomogeneous inner solution
to give the steady hydrodynamic force obtained by \citet{wang2012inertial} and \citet{khair2014expansions}. 

Conversely, when unsteadiness is high,  $\ell_p / \ell_{O}  \to 0$. In this limit $A(t,\:\tau) \to 0$. As a result,
 the integrals converge to
        \begin{equation}
        \boldsymbol{I}_1 \to - \varepsilon\int_0^t \rm{d} \tau\frac{\ma{I}}{\sqrt{\pi(t-\tau)}}\cdot \frac{\rm{d} \vec{f}^{(0)}(\tau)}{\rm{d} \tau}\,,\quad \mbox{and} \quad
        \boldsymbol{I}_2 \to \vec{0}\:,
        \label{eq:to_unsteadiness}
        \end{equation}
where $\ma{I}$ is the identity tensor. 
So the kernel $\ma{K}^{(1)}$ in $\vec I_1$ simplifies to the  BBO  kernel \citep{boussinesq1885resistance,basset1888treatise} which describes how the disturbance-velocity gradients relax due to viscous diffusion.   In this limit, the $\varepsilon$-terms dominate over the $\rm{Re}_{\rm{p}}$-terms, and  the outer flow (\ref{W_outer}) converges to the outer flow  obtained from the unsteady Stokes equation. This is the limit considered by \cite{wang2012unsteady} and \cite{redaelli2021}. The fact that the BBO history force is contained in Eq. (\ref{force_re}) shows that
the perturbation theory is uniformly valid in $\varepsilon$  \citep[as in][]{mehaddi2018inertial}. In other words, Eq.~(\ref{force_re}) is accurate if the particle Reynolds number is small, regardless of how large the unsteadiness of the problem.  
Note, however, that when the unsteadiness is very large, for $\varepsilon\gg 1$, there is an additional contribution 
to the hydrodynamic force: the added-mass force  \citep{wang2012unsteady,candelier2023second} which is of order $\varepsilon^2$.

In summary, there are two fundamental differences between the hydrodynamic force on an active  compared to a passive sphere in a time-dependent uniform flow.  First, while the history kernels are likely to be the same, the other factors in the integrals are different because they involve an active part for the active particle. Second, for an active particle there are additional instantaneous contributions to the hydrodynamic force that stem from an inhomogeneous solution of the inner problem. Spherical symmetry causes these contributions to vanish for the passive sphere. This does not happen for the spherical swimmer, because swimming breaks the spherical  symmetry.

\subsection{Effect of convective and unsteady fluid inertia on the dynamics}

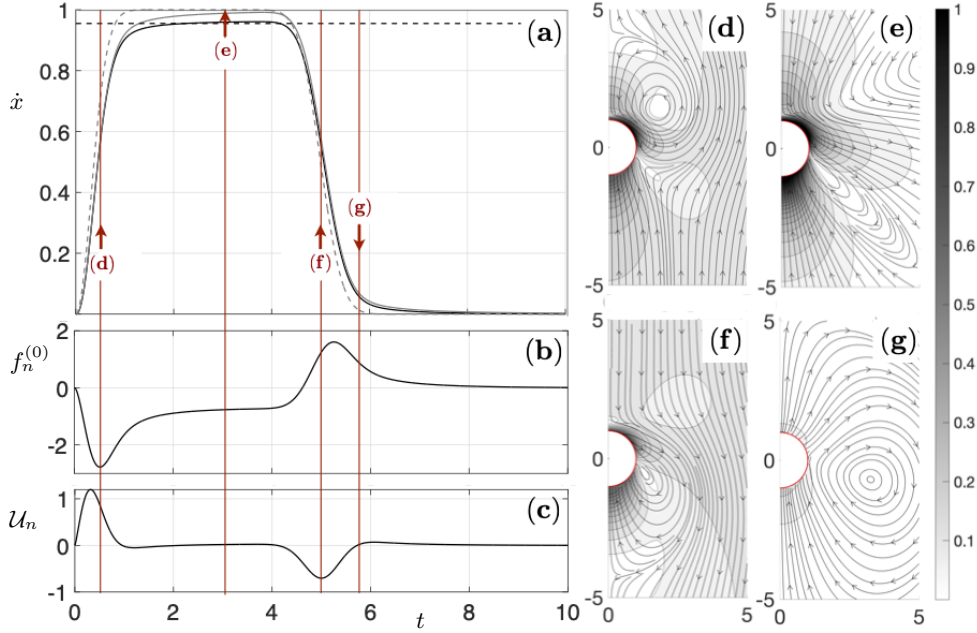
\begin{figure}
\centering
\begin{overpic}[width=15cm]{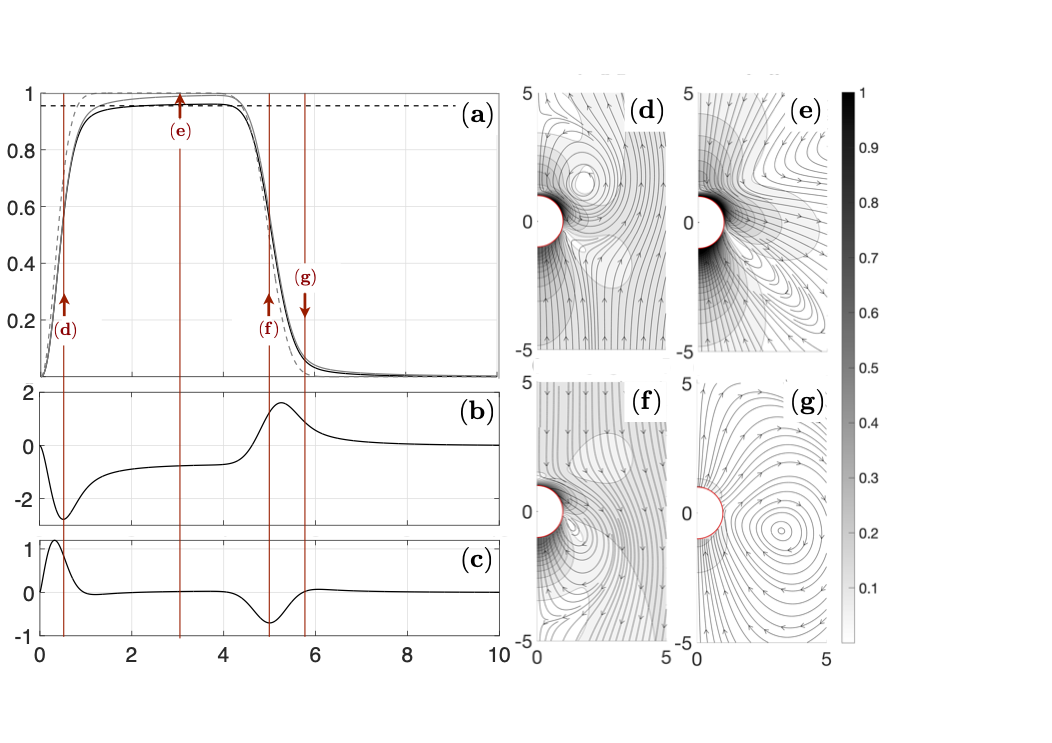}
\put(-2,53){$\dot{x}$}
\put(-2,30){$f^{(0)}_n$}
\put(-2,16){$\mathcal{U}_n$}
\put(34,7){$t$}
\end{overpic}
\caption{\label{fig:velocity}  {\bf Sudden start followed by sudden stop} [Eq.~(\ref{eq:sudden_start})]. (a) Time dependence of the resulting centre-of-mass speed $\dot x$ non-dimensionalised with $u_c$
and $t_\nu$ (solid black line). Also shown is the BBO approximation obtained using Eq.~({2.12}) of   \citet{wang2012unsteady} (solid gray line). The dashed gray line shows
the Stokes approximation for the centre-of-mass speed. 
 The horizontal dashed black line corresponds to  the 
steady-state speed~\citep{wang2012inertial,khair2014expansions}.  (b) Time dependence of  $\vec{f}^{(0)} \cdot \vec{n}$
[see Eq.~(\ref{eq:f0_def})].  (c) Time dependence of  $\vec{\mathcal{U}}\cdot \vec{n}$. 
Panels (d)--(g) show the  disturbance flow produced by the squirmer at the different non-dimensional times  $t=0.5,\: 3,\: 5$, and $5.6$,
the streamlines as well as the magnitude. 
 }
\end{figure}

As an example, consider the motion a small neutrally buoyant swimmer that jumps. Its motion starts suddenly, followed by a sudden stop.
We mimic the sudden start/stop by imposing a corresponding time dependence on the  tangential surface velocity:
\begin{equation}
\label{eq:sudden_start}
B_1(t) = \left[ 1 - \mbox{erf}\left( {2t}/{\tau_\nu}- 10 \right)\right]\mbox{erf}\left({2t}/{\tau_\nu}\right)^2 \quad \mbox{(mm/s)} \quad \mbox{and} \quad B_2(t)=\tfrac{3}{2} B_1(t)
\end{equation}
in dimensional variables.  The error function squared in Eq.~(\ref{eq:sudden_start}) ensures  that $\vec f^{(0)}$ and its time derivative both vanish at $t=0$,  as required by Eq.~(\ref{force_re}). 

Consider how to non-dimensionalise the problem. The maximal value of $B_1(t)$ produces a swimming speed of  $ \tfrac{4}{3} $ mm/s
in the Stokes limit, and we chose
to non-dimensionalise with $u_c=\tfrac{4}{3}$ mm/s.
For particle of radius $a = 150$ $\mu$m in water ($\nu = 10^{-6}$ m$^2$/s), this gives  ${\rm Re}_{\rm{p}}$=0.2. 
How to choose the time scale $\tau_c$ is perhaps less obvious.  
The  typical time of variation of the boundary conditions on the surface of the swimmer can be  estimated by
$B_1(t)/\dot{B}_1(t)$.
 At short times, $B_1(t)/\dot{B}_1(t) \approx  \tfrac{1}{2} t$. This shows that the characteristic time tends to zero at very short times.
After a while, however, the parameter $B_1(t)$ reaches a plateau, the disturbance flow becomes steady, and the characteristic time tends to infinity. 
As discussed at the end of \S \ref{section_Hydro_force}, Eq.~(\ref{force_re}) is valid regardless of how unsteady the motion is. It is only required
that the particle Reynolds number is small. As a consequence, it does not matter precisely how the time scale $\tau_c$ is chosen. We took $\tau_c= \tau_\nu $.  

To determine the trajectory, one needs to solve
Newton's equation of motion. In non-dimensional variables it reads
\begin{equation}
\tfrac{4 \pi}{3} \varepsilon^2 \ddot{{\vec x}} = \vec f'  \:.
\label{Eq:mvt}
\end{equation}
Both fluid and the particle inertia are accounted for in $\vec f^\prime$. Considering the expression for the hydrodynamic force,  Eq. (\ref{force_re}), we see
that  Eq.~(\ref{Eq:mvt})  is  an integro-differential equation. We solved this equation numerically using the method described by \cite{daitche2013advection}.
When only unsteady fluid inertia is considered, one can solve the equation of motion by Laplace transform \citep{Ishimoto2013,fouxon2019inertial,wang2012unsteady}.
In our case this is not possible, because the equation is non-linear.

Figure~\ref{fig:velocity} illustrates how dynamics and the disturbance flow develop after a sudden start followed by a sudden stop, 
obtained by solving Eqs.~(\ref{force_re},\ref{Eq:mvt}) together with (\ref{eq:sudden_start}).
Panel (a)  shows how the centre-of-mass speed $\dot x$ changes when both unsteady and convective fluid inertia are taken into account (solid  black line). At very short times, the centre-of-mass motion of the  squirmer
 is  well described by the BBO equation which neglects the effect of convective fluid inertia (solid gray  line). This is expected, because the dynamics is dominated
 by the unsteadiness of the disturbance flow at short times. Also shown in panel (a) is the Stokes swimming speed (2/3) $B_1(t)$ (dashed gray line). We see that the actual swimming
 velocity differs significantly from the Stokes approximation: 
  during acceleration phase,  $\dot{x}$ is lower than 
$(2/3)\:B_1$. This produces a negative value of the component of  the force  $\vec{f}^{(0)}$ along the swimming direction $\vec{n}$ [panel (b)],
 and this affects 
the uniform contribution $\vec{\mathcal{U}}$  [panel (c)].

At larger times,  during the plateau of $B_1(t)$, differences between the short-time BBO approximation and the present theory appear, caused by 
 the instantaneous contributions from the inner solution to the hydrodynamic force [the last  two terms on the r.h.s. of Eq.~(\ref{force_re})]. 
Although the inertial parameters are quite small (as they must be for the theory to be valid), these terms nevertheless have a noticeable effect upon the centre-of-mass speed.
\citet{khair2014expansions} and \citet{wang2012inertial} derived an approximation for the centre-of-mass speed in the steady limit, where $B_1$ and $B_2$ are independent of time,
\begin{equation}
\dot{\vec x} \approx \tfrac{2}{3}B_1(1 - \tfrac{3 \beta}{20}\mathrm{Re}_{\rm{p}} ) \vec{n}\,.
\end{equation}
 This result is shown as a horizontal dashed black line in panel (a). Comparing with our theory, we observe that $\dot x$ reaches its steady limit after a few viscous times. 
 As a consequence, the magnitude of the hydrodynamic force  decreases   to  $ \vec{f}^{(0)} \approx-\tfrac{6 \pi }{10} B_2 \mathrm{Re}_{\rm{p}}  \vec{n}$ (where $B_2 = \beta B_1$ was used), 
  and  $\vec{\mathcal{U}}$  decreases to  $\tfrac{1}{40} B_1 B_2 \:\mathrm{Re}_{\rm{p}}^2\vec n$. Comparing the two expressions, we see that
  convective-inertia effects due to the singular  term $\vec{\mathcal{U}}$ are negligible at small Re$_p$, because it only contributes at order  $\mathcal{O}(\mathrm{Re}_{\rm{p}}^2)$.
Now consider the deceleration phase.  As the parameter $B_1 \to 0$,  the Stokes  velocity decreases to zero. When inertia effects are taken into account, however, $\dot x$ relaxes more slowly. 
The present theory and the BBO approximation yield quite similar results in this phase. This suggests that the decrease of the velocity of the organism for a sudden stop is mainly governed by unsteady-inertia effects.

Panels (d)--(g) illustrate how the disturbance near the swimmer changes as a function of time. Shown are streamlines and contours of the magnitude
 $|{\vec{w}}_{\mbox{\scriptsize in}}^{(0)} + \varepsilon  {\vec{w}}_{\mbox{\scriptsize in}}^{(1)}|$
of the disturbance-flow velocity  
at four  different  times, marked by arrows in panel (a). Panel~(d) shows the flow during the acceleration phase. 
The term $\vec{\mathcal{U}}\cdot \vec{n}$  contributes a Stokeslet to the disturbance flow  which decays slowly, as $r^{-1}$, and thus dominates far from the swimmer. As a consequence, the disturbance flow extends far from the swimmer. Near its surface, on the other hand, the disturbance must match the imposed tangential surface velocity; in the upper half plane this velocity opposes the swimming velocity. Therefore a  growing flow cell forms in front of the swimmer.
Panel (e) shows the disturbance at a later stage, when the motion is quasi steady, so that  $\vec{\mathcal{U}}$ vanishes. The disturbance flow is
essentially a stresslet that decays  as $r^{-2}$. The disturbance is therefore localised closer to the swimmer. We checked that the flow field is well approximated by the steady solution obtained by~\cite{Chisholm2016}.
 In panel (f) we plot the disturbance during the deceleration phase. This case is similar to that shown in panel (d), except that  the sign of  $\vec{\mathcal{U}}$ is reversed.
Near the swimmer, the flow cell now forms behind the swimmer. 
 Panel (g) shows the disturbance flow at large times. Here, the parameter $B_1$ is almost zero. The disturbance flow is therefore
 similar to that produced by  a passive sphere, a Stokeslet, but with a very weak intensity.

 \section{Conclusions}
\label{sec:conc}

We determined the hydrodynamic force on a small spherical squirmer in an unsteady, spatially homogeneous flow. We obtained the hydrodynamic force
by asymptotic matching of perturbation expansions in the parameter $\varepsilon = \sqrt{{\rm Re}_{\rm{p}} {\rm Sl}}$, for small particle Reynolds numbers. 
Our main result (\ref{force_re}) for the hydrodynamic force generalises the result of \citet{lovalenti1993hydrodynamic} from a passive sphere to an active particle, an unsteady spherical squirmer. Equation~(\ref{force_re}) describes how convective inertia changes the kernel of the history force. As explained by \citet{lovalenti1993hydrodynamic}, convective
inertia tends to cause the kernel to decay more rapidly. We believe that the kernels for passive and active particles are identical. We did  not demonstrate  this analytically, but numerical evaluation for all cases we considered showed them to be the same. The kernels do not depend upon the particular swimming gait of the squirmer,  given by the coefficients $B_1(t)$ and $B_2(t)$. This information is encoded in $\vec f^{(0)}$.  
For the active particle, this amplitude contains an additional term, compared with the passive sphere, stemming from the active surface-velocity field.

A second difference to the result of \citet{lovalenti1993hydrodynamic} is that an inhomogeneous part of the inner solution contributes to the hydrodynamic force. Spherical symmetry ensures that this contribution vanishes for a spherical passive particle, but swimming breaks spherical symmetry. This contribution explains why the  method used by \cite{redaelli2021}
to compute inertial corrections to the hydrodynamic force works in the Saffman limit and for unsteady inertia, but not for the Oseen problem considered here. 

Our expression (\ref{force_re}) for the hydrodynamic force simplifies to known results in two limits. First, when unsteadiness dominates, our result
simplifies to that of  \citet{wang2012unsteady} and \citet{redaelli2021}, where the history force is determined by the BBO kernel which decays as $t^{-1/2}$. 
Second, when particle inertia is more important than unsteadiness, our expression converges to  the steady Oseen approximation obtained by \citet{wang2012inertial} and \citet{khair2014expansions}.

In order to illustrate the effects of weak unsteady and convective fluid inertia, we considered  a swimmer 
that suddenly starts its centre-of-mass motion, followed by a sudden stop. 
We analysed how the disturbance flow created by the swimmer changes as a function of time. 
During  acceleration and deceleration, the disturbance flow  is essentially a Stokeslet which decreases more slowly far from
the swimmer than the stresslet due to steady swimming.  This implies that the swimmer is easier to detect  immediately after a sudden start or a sudden stop, because the disturbance  can be perceived from further away.

We stress that the theory presented here rests on the assumption that Re$_{\rm{p}} \ll 1$. Marine organisms come in many different sizes, and they swim with different speeds and with different swimming gaits. A number of empirical studies have estimated both the particle Reynolds number Re$_{\rm{p}}$ and the Strouhal number $\rm{Sl}$ for different microswimmers. \citet{Wadhwa2014}  estimated the Strouhal number for copepods,
and concluded that ${\rm Sl} \sim 1$ for nauplii (with Re$_{\rm p} = 5\ldots 10$) and for adult copepods  that move more vigorously (Re$_{\rm{p}} \sim 40$). 
\cite{kiorboe2014} measured Re$_{\rm{p}}$ and $\rm Sl$ for copepods in different stages of their evolution, observing $\rm Sl \sim 1$, and Re$_{\rm{p}}$-values up to $70$  (see their Tab.~S1). For these values of Re$_{\rm{p}}$, our theory most certainly fails. 
However, there are also organisms that move less vigorously.  {\em Mesodinium rubrum}, for instance, has particle Reynolds numbers of order of unity \citep{Jiang2011Robidium}. Also, cruising copepods tend to swim with smaller particle Reynolds numbers, in the range $\sim 0.1 - 0.4$, see Tables I and II in \citep{qiu2022gyrotactic}. In these cases, small-Re$_p$ theory may  give a qualitatively correct description of the dynamics.  

Let us consider in more detail how fluid inertia affects  the dynamics of a small  neutrally buoyant organism that use the ciliary propulsion to swim. For such organisms, the instantaneous velocity produced by the oscillations of the cilia over the  surface of the particle scales as $\dot{x} \sim \epsilon_s a \omega$, where  $\epsilon_s a$ is  the amplitude of the oscillation of the cilia ($\epsilon_s\sim 0.1$ is a small non-dimensional parameter), and $\omega$ is the angular frequency of the oscillations. 
It follows that  ${\rm{Re}}_{\rm{p}} = \epsilon_s a^2 \omega/\nu = \epsilon_s \varepsilon^2$.  So when Re$_p$ is unity, unsteady inertia
dominates the dynamics, as for instance for {\it Paramecium}. In this case, Eq.~(\ref{force_re})  shows that the hydrodynamic force is well approximated using
the BBO kernel, and adding the instantaneous convective-inertia contribution  mentioned above.

When the swimmer is not neutrally buoyant, the particle equation-of-motion contains an additional term, the  buoyancy force. 
In this case, the amplitude  $\vec{f}^{(0)}$ does not tend  to zero, even if the swimmer stops swimming,  because $\vec{f}^{(0)}$ must balance the buoyancy force. 
To leading order in Re$_p$, the resulting convective-inertia correction is described by our result for the hydrodynamic force.

The disturbance caused by small motile organisms in a marine environment has been described in other ways, by superposing different elementary Stokes solutions. Examples are the impulsive Stokeslet 
and the impulsive stresslet \citep[][]{afanasyev2004wakes}. \citet{guasto2012fluid} give examples where this approach fails, because it does not reliably approximate the outer disturbance flow,
and they state possible reasons: inertia effect induced by the unsteadiness, buoyancy 
\citep[giving rise, for example, to a Stratelets,][]{ardekani2010stratlets}, 
or a combination of both.
At least for small particle Reynolds numbers, our results show how convective and unsteady fluid inertia modify the disturbance flow.
At small Reynolds number, the outer flow is universal: shape and swimming gait enter only through an amplitude, the kernels describing the history effect on the outer disturbance flow do not depend on these details.
However, for larger particle Reynolds numbers it remains an open question how history force and disturbance flow depend on the shape and propulsion mechanism of the swimmer.

	\vfill\eject

\section*{Declaration of Interests} 
The authors report no conflict of interest.

	\section*{Acknowledgments}  
	The research of BM was supported by  Vetenskapsr\aa{}det, grant no. 2021-4452, and in part also by the Knut and Alice Wallenberg Foundation, grant no. 2019.0079.  
	TR and CE were supported by  funding from the European Research Council (ERC) under the European Union's Horizon 2020 research and innovation programme (grant agreement no. 834238).

\appendix
 \label{AppendixA}	
\section{}
\renewcommand{\theequation}{A\arabic{equation}}

In this Appendix we summarise how the outer solution  (\ref{eq:outer_solution}) of Eqs.~(\ref{eq:eom52}) is obtained. Since Eq.~(\ref{eq:outer_a}) contains
a source term with a Dirac delta function  $\delta(\vec{r})$, it is most conveniently solved by Fourier transform.
We use the convention
	\begin{equation}
	\hat{\vec{w}} = \int \mbox{d}^3 \vec{r} \,\,  \vec{w}\exp(- i \vec{k}\cdot\vec{r})
	\quad \mbox{and} \quad 
	\vec{w} =  \frac{1}{8\pi^3}\int \mbox{d}^3 \vec{k} \:\hat{\vec{w}} \exp( i \vec{k}\cdot\vec{r}) \:.
	\end{equation}
	Transforming Eqs.~(\ref{eq:eom52}) one finds: 
	\begin{subequations}
	\begin{align}
	&\vec{k} \cdot \hat{\vec{w}}_{\mbox{\scriptsize out}}  =0\:,	\label{eq:divergence_fourier}\\
	&
	\varepsilon^2 \frac{\partial  \hat{\vec{w}}_{\mbox{\scriptsize out}}}{\partial t} -  i \varepsilon  \left(\frac{\ell_p}{\ell_{O}} \right)  (\dot{\vec x} \cdot\vec{ k}) \: \hat{\vec{w}}_{\mbox{\scriptsize out}} 
	= - i \vec{ k} \: \hat{p}_{\mbox{\scriptsize out}}  - k^2  \hat{\vec{w}}_{\mbox{\scriptsize out}}  + \vec{f}^{(0)} \:.
	\label{eq:far_field2}
	\end{align}
	\end{subequations}
	The pressure $\hat{p}_{\mbox{\scriptsize out}}$ can be determined by projecting Eq.~(\ref{eq:far_field2}) along $\vec{k}$, and using Eq.~(\ref{eq:divergence_fourier}).  Substituting 
	the resulting expression for $\hat{p}_{\mbox{\scriptsize out}}$  into Eq.(\ref{eq:far_field2}) yields 
	\begin{equation}
	\varepsilon^2  \frac{\partial  \hat{\vec{w}}_{\mbox{\scriptsize out}}}{\partial t} =  - k^2  \hat{\vec{w}}_{\mbox{\scriptsize out}}  +  i \varepsilon  \left(\frac{\ell_p}{\ell_{O}} \right)  (\dot{\vec x}\cdot\vec{ k})  \hat{\vec{w}}_{\mbox{\scriptsize out}} 
	+ k^2 \:\hat{\ma {G}} \cdot \vec{f}^{(0)} \:,
	\label{eq:far_field3}
	\end{equation}
	where $\hat{\ma G}$ is the  Fourier transform of the Green tensor of the Stokes equations,
	\begin{align}
	 [ \hat{\ma G}]_{i j} (\vec k)= \frac{1}{k^2}\left( \delta_{ij} - \frac{k_ik_j}{k^2} \right)\:,\quad
	 [\ma{G}]_{i j} (\vec r)= \frac{1}{8 \pi} \left(\frac{\delta_{i j}}{r} + \frac{r_i r_j}{r^3} \right)\,.
	\end{align}
	The solution of the inhomogeneous differential equation~(\ref{eq:far_field3}) reads
	\begin{equation}
	\hat{\vec{w}}_{\mbox{\scriptsize out}} = \frac{k^2}{\varepsilon^2}  \int_0^t\mbox{d}\tau 
	\: \:\hat{\ma {G}} \cdot \vec{f}^{(0)}(\tau) 
	\exp\Bigg[-\frac{k^2}{\varepsilon^2} (t-\tau) 
	+ \frac{i}{\varepsilon}\left(\frac{\ell_p}{\ell_{O}}\right)  \int_{\tau}^t \mbox{d}  \tau'\, \vec{ k}\cdot \dot{\vec x} (\tau') 
	\Bigg] \:.
	\label{Eq_outer1}
	\end{equation}
	Here we assumed that the force $\vec{f}^{(0)}$ vanishes for $t\leq 0$.
	
	In order to match the outer solution  (\ref{Eq_outer1})  to the inner solution, we
 seek an expansion of the outer solution  of the form
	\begin{equation}
	\hat{{\vec{w}}}_{\mbox{\scriptsize out}}  = \hat{\mathcal{T}}_{\mbox{\scriptsize reg}}^{(0)}  + \varepsilon \hat{\mathcal{T}}_{\mbox{\scriptsize reg}}^{(1)} +\varepsilon \hat{\mathcal{T}}_{\mbox{\scriptsize sing}}^{(1)} + \mathcal{O}(\varepsilon^2)\:.
	\label{W_outer1}
	\end{equation}
	The terms $\hat{\mathcal{T}}_{\mbox{\scriptsize reg}}^{(n)}$ correspond to regular parts of the expansion.
	From Eq. (\ref{eq:far_field3})  we are led to
	\begin{eqnarray}
	\widehat{\mathcal{T}}_{\mbox{\scriptsize reg}}^{(0)}  &= &\hat{\ma{G}} \cdot  \vec{f}^{(0)}  
	\label{T0reg}\:,\\
	\widehat{\mathcal{T}}_{\mbox{\scriptsize reg}}^{(1)}  &=& i \left(\frac{\ell_p}{\ell_{O}} \right)  (\dot{\vec x} \cdot\vec{k})\frac{ \hat{\ma{G}}\cdot  \vec{f}^{(0)} }{k^2}   
	\label{T1reg}\:. 
	\end{eqnarray}
	Transforming back to configuration space gives 	
	\begin{eqnarray}
	{\mathcal{T}}_{\mbox{\scriptsize reg}}^{(0)}  &= &{\ma{G}} \cdot  \vec{f}^{(0)}  
	\label{T0regphys}\:,\\
	{\mathcal{T}}_{\mbox{\scriptsize reg}}^{(1)}  &=& - \left(\frac{\ell_p}{\ell_{O}}\right)\dot{\vec x}\cdot \boldsymbol{\nabla}\left[\frac{3 r}{32\pi} \left( \ma{1} - \frac{1}{3} \frac{\vec{r} \otimes \vec{r}}{r^2}\right) \cdot \vec{f}^{(0)} \right]\,.
	\label{T1regphy}
	\end{eqnarray}
	The term $\hat{\mathcal{T}}_{\mbox{\scriptsize sing}}^{(1)}$ is  singular in $\vec k$-space \citep{Meibohm2016}, proportional to $\delta(\vec k)$.
The singular term in the expansion (\ref{W_outer1}) is determined by evaluating the limit \citep{Childress64,saffman1965lift,Meibohm2016}
	\begin{equation}
	\widehat{\mathcal{T}}_{\mbox{\scriptsize sing}}^{(1)}   =  \lim_{\varepsilon \to 0} \frac{\widehat{{\vec{w}}}_{\mbox{\scriptsize out}}  - \widehat{\mathcal{T}}_{\mbox{\scriptsize reg}}^{(0)}}{\varepsilon} - \widehat{\mathcal{T}}_{\mbox{\scriptsize reg}}^{(1)} \:.
	\label{eqT1sing}
	\end{equation}
	Evaluating the limit, one finds 
		\begin{equation}
	\widehat{\mathcal{T}}_{\mbox{\scriptsize sing}}^{(1)}   = 8 \pi^3 \vec{\mathcal{U}}(t)\: \delta(\vec{k})\:,
	\label{eq:Ai}
	\end{equation}
	with
	\begin{align}
	\nonumber
	\vec{\mathcal{U}}(t) = &  \frac{1}{8\pi^3}\int \mbox{d}^3\vec{k} \left\{ - \int_0^t\!\!\mbox{d}\tau\, \hat{\ma{G}} \cdot \frac{\mbox{d} \vec{f}^{(0)}(\tau)}{\mbox{d} \tau} \exp\Big( -k^2 (t-\tau) + 2 i \sqrt{t-\tau }\, \vec{ k}\cdot\vec{a}(t,\tau) \Big)  \right.\\
	& \left.\hspace*{-7mm}
	+  i
	\left(\frac{\ell_p}{\ell_{O}}\right) 
	\int_0^t \!\!\mbox{d} \tau\, \hat{\ma{G}}  \cdot \vec{f}^{(0)}(\tau)(\vec{k}\cdot\dot{\vec{x}}(\tau))
	\exp\Big( -k^2 (t\!-\!\tau) + 2 i \sqrt{t\!-\!\tau }\, \vec{k}\cdot\vec{a}(t,\tau) \Big) \right\}\:.
	\label{eq:force}
	\end{align}
	Here, $\vec{a}$  is given in Eq. (\ref{eq:a}). 
	In order to perform the $\vec{k}$-integration in Eq.~(\ref{eq:force}), we follow \cite{lovalenti1993hydrodynamic} and express the vectors in spherical coordinates defined around an axis that moves with the displacement vector $\vec{a}$.   We therefore write
	\begin{equation}
	\label{eq:pdef}
	\vec{k}\cdot \vec{a}  = k A(t,\:\tau)\cos(\theta_a)\quad \mbox{and} \quad \hat{\vec{q}} = \frac{\vec{a}(t,\:\tau)}{A(t,\tau)}
	\end{equation}
	where $\theta_a$ is the angle between $\vec{k}$ and $\vec{a}$, and $A(t,\:\tau)$ is the norm of $\vec{a}$. 
	Performing the $\vec{k}$-integration yields
	\begin{equation}
	\vec{\mathcal{U}}(t)  = -\frac{1}{6\pi}  \int_{0}^{t}\mbox{d}\tau \, {\ma{K}^{(1)}(t,\tau) }\cdot \frac{{\mbox{ d}}\vec{f}^{(0)}(\tau)}{\mbox{ d} \tau}   -   \left(\frac{\ell_p}{\ell_{O}}\right) \frac{1}{6\pi}  \int_{0}^{t} \mbox{d}\tau\,{\ma{ K}^{(2)}(t,\tau)}  \cdot \vec{f}^{(0)}(\tau)\,.
	\label{eq:F_disturb}
	\end{equation}
	The expressions for the kernels $\ma{K}^{(1)}$ and $\ma{K}^{(2)}$ are given in Eqs.~(\ref{eq:K1}) and (\ref{eq:K2}).


\begin{thebibliography}{42}
\expandafter\ifx\csname natexlab\endcsname\relax\def\natexlab#1{#1}\fi

\bibitem[Afanasyev(2004)]{afanasyev2004wakes}
{\sc Afanasyev, Yakov} 2004 Wakes behind towed and self-propelled bodies:
  Asymptotic theory. {\em Physics of Fluids\/} {\bf 16}, 3235--3238.

\bibitem[Ardekani \& Stocker(2010)]{ardekani2010stratlets}
{\sc Ardekani, A.~M. \& Stocker, R.} 2010 Stratlets: Low {R}eynolds number
  point-force solutions in a stratified fluid. {\em Phys. Rev. Lett.\/} {\bf
  105}, 084502.

\bibitem[Basset(1888)]{basset1888treatise}
{\sc Basset, A.~B.} 1888 {\em A treatise on hydrodynamics: with numerous
  examples.\/}, , vol.~2. Deighton, Bell and Company.

\bibitem[Blake(1971)]{Blake1971}
{\sc Blake, J.~R.} 1971 A spherical envelope approach to ciliary propulsion.
  {\em J. Fluid Mech.\/} {\bf 46}, 199--208.

\bibitem[Boussinesq(1885)]{boussinesq1885resistance}
{\sc Boussinesq, J.} 1885 Sur la resistance qu'oppose un fluide indefini en
  repos, sans pesanteur, au mouvement varie d'une sphere solide qu'il mouille
  sur toute sa surface, quand les vitesses restent bien continues et assez
  faibles pour que leurs carres et produits soient negligeables. {\em CR Acad.
  Sc. Paris\/} {\bf 100}, 935--937.

\bibitem[Candelier {\em et~al.\/}(2023)Candelier, Mehaddi, Mehlig \&
  Magnaudet]{candelier2023second}
{\sc Candelier, F., Mehaddi, R., Mehlig, B. \& Magnaudet, J.} 2023 Second-order
  inertial forces and torques on a sphere in a viscous steady linear flow. {\em
  Journal of Fluid Mechanics\/} {\bf 954}, A25.

\bibitem[Childress(1964)]{Childress64}
{\sc Childress, S.} 1964 The slow motion of a sphere in a rotating, viscous
  fluid. {\em J. Fluid Mech.\/} {\bf 20}~(2), 305--314.

\bibitem[Chisholm {\em et~al.\/}(2016)Chisholm, Legendre, Lauga \&
  Khair]{Chisholm2016}
{\sc Chisholm, N.~G., Legendre, D., Lauga, E. \& Khair, A.~S.} 2016 A squirmer
  across {R}eynolds numbers. {\em J. Fluid Mech.\/} {\bf 796}, 233--256.

\bibitem[Daitche(2013)]{daitche2013advection}
{\sc Daitche, Anton} 2013 Advection of inertial particles in the presence of
  the history force: Higher order numerical schemes. {\em Journal of
  Computational Physics\/} {\bf 254}, 93--106.

\bibitem[Fauci \& Dillon(2006)]{fauci2006}
{\sc Fauci, L.~J. \& Dillon, R.} 2006 Biofluidmechanics of reproduction. {\em
  Annu. Rev. Fluid Mech.\/} {\bf 38}~(1), 371--394.

\bibitem[Fenchel \& Juel~Hansen(2006)]{fenchel2006motile}
{\sc Fenchel, Tom \& Juel~Hansen, Per} 2006 Motile behaviour of the
  bloom-forming ciliate mesodinium rubrum. {\em Marine Biology Research\/} {\bf
  2}~(01), 33--40.

\bibitem[Fouxon \& Or(2019)]{fouxon2019inertial}
{\sc Fouxon, Itzhak \& Or, Yizhar} 2019 Inertial self-propulsion of spherical
  microswimmers by rotation-translation coupling. {\em Physical Review
  Fluids\/} {\bf 4}~(2), 023101.

\bibitem[Guasto {\em et~al.\/}(2012)Guasto, Rusconi \&
  Stocker]{guasto2012fluid}
{\sc Guasto, Jeffrey~S., Rusconi, Roberto \& Stocker, Roman} 2012 Fluid
  mechanics of planktonic microorganisms. {\em Annual Review of Fluid
  Mechanics\/} {\bf 44}~(1), 373--400.

\bibitem[Happel \& Brenner(1965)]{happelbrenner1981}
{\sc Happel, J. \& Brenner, H.} 1965 {\em Low {Reynolds} number hydrodynamics:
  with special applications to particulate media\/}. Prentice-Hall.

\bibitem[Hinch(1995)]{Hinch1995}
{\sc Hinch, E.~J.} 1995 {\em Perturbation Methods.\/}. Cambridge University
  Press.

\bibitem[Ishimoto(2013)]{Ishimoto2013}
{\sc Ishimoto, K.} 2013 A spherical squirming swimmer in unsteady stokes flow.
  {\em J. Fluid Mech.\/} {\bf 723}, 163--189.

\bibitem[Jiang(2011)]{Jiang2011Robidium}
{\sc Jiang, H.} 2011 Why does the jumping ciliate \textit{Mesodinium rubrum}
  possess an equatorially located propulsive ciliary belt? {\em J. Plank.
  Res.\/} {\bf 33}~(7), 998--1011.

\bibitem[Jiang \& Kiorboe(2011)]{Jiang2011}
{\sc Jiang, H. \& Kiorboe, T.} 2011 {The fluid dynamics of swimming by jumping
  in copepods}. {\em J. R. Soc. Interface\/} {\bf 8}~(61), 1090--1103.

\bibitem[Khair \& Chisholm(2014)]{khair2014expansions}
{\sc Khair, A.~S. \& Chisholm, N.~G.} 2014 Expansions at small reynolds numbers
  for the locomotion of a spherical squirmer. {\em Phys. Fluids\/} {\bf
  26}~(1), 011902.

\bibitem[Kiorboe {\em et~al.\/}(2014)Kiorboe, Jiang, Goncalves, Nielsen \&
  N.]{kiorboe2014}
{\sc Kiorboe, T., Jiang, H., Goncalves, R.~J., Nielsen, L.~T. \& N., Wadhwa}
  2014 Flow disturbances generated by feeding and swimming zooplankton. {\em
  PNAS\/} pp. 11738--11743.

\bibitem[Lauga \& Powers(2009)]{lauga2009}
{\sc Lauga, E. \& Powers, T.~R.} 2009 The hydrodynamics of swimming
  microorganisms. {\em Rep. Prog. Phys.\/} {\bf 72}~(9), 096601, arXiv:
  0812.2887.

\bibitem[Legendre \& Magnaudet(1998)]{legendre1998lift}
{\sc Legendre, D. \& Magnaudet, J.} 1998 The lift force on a spherical bubble
  in a viscous linear shear flow. {\em J. Fluid Mech.\/} {\bf 368}, 81--126.

\bibitem[Lighthill(1952)]{Lighthill1952}
{\sc Lighthill, M.~J.} 1952 {On the squirming motion of nearly spherical
  deformable bodies through liquids at very small reynolds numbers}. {\em
  Communications on Pure and Applied Mathematics\/} {\bf 5}~(2), 109--118.

\bibitem[Lovalenti \& Brady(1993)]{lovalenti1993hydrodynamic}
{\sc Lovalenti, P.~M. \& Brady, J.~F.} 1993 The hydrodynamic force on a rigid
  particle undergoing arbitrary time-dependent motion at small reynolds number.
  {\em J. Fluid Mech.\/} {\bf 256}, 561--605.

\bibitem[Mehaddi {\em et~al.\/}(2018)Mehaddi, Candelier \&
  Mehlig]{mehaddi2018inertial}
{\sc Mehaddi, R., Candelier, F. \& Mehlig, B.} 2018 Inertial drag on a sphere
  settling in a stratified fluid. {\em J. Fluid Mech.\/} {\bf 855}, 1074--1087.

\bibitem[Meibohm {\em et~al.\/}(2016)Meibohm, Candelier, Ros\'en, Einarsson,
  Lundell \& Mehlig]{Meibohm2016}
{\sc Meibohm, J., Candelier, F., Ros\'en, T., Einarsson, J., Lundell, F. \&
  Mehlig, B.} 2016 Angular velocity of a spheroid log rolling in a simple shear
  at small {Reynolds} number. {\em Phys. Rev. Fluids\/} {\bf 1}.

\bibitem[Oseen(1927)]{oseen1927neuere}
{\sc Oseen, C.~W.} 1927 Neuere {M}ethoden und {E}rgebnisse in der
  {H}ydrodynamik. {\em Leipzig: Akademische Verlagsgesellschaft m. b. H.\/} .

\bibitem[Pak \& Lauga(2014)]{Pak2014}
{\sc Pak, O.~S. \& Lauga, E.} 2014 Generalized squirming motion of a sphere.
  {\em J. Eng. Math.\/} {\bf 88}, 1--28.

\bibitem[Pedley(2016)]{pedleysquirmer}
{\sc Pedley, T.~J.} 2016 Spherical squirmers: models for swimming
  micro-organisms. {\em IMA J. Appl. Math.\/} {\bf 81}, 488–521.

\bibitem[Qiu {\em et~al.\/}(2022)Qiu, Cui, Climent \& Zhao]{qiu2022gyrotactic}
{\sc Qiu, Jingran, Cui, Zhiwen, Climent, Eric \& Zhao, Lihao} 2022 Gyrotactic
  mechanism induced by fluid inertial torque for settling elongated
  microswimmers. {\em Phys. Rev. Research\/} {\bf 4}, 023094.

\bibitem[Redaelli {\em et~al.\/}(2022)Redaelli, Candelier, Mehaddi \&
  Mehlig]{redaelli2021}
{\sc Redaelli, T., Candelier, F., Mehaddi, R. \& Mehlig, B.} 2022 Unsteady and
  inertial dynamics of a small active particle in a fluid. {\em Phys. Rev.
  Fluids\/} {\bf 7}, 044304.

\bibitem[Saffman(1965)]{saffman1965lift}
{\sc Saffman, P. G.~T.} 1965 The lift on a small sphere in a slow shear flow.
  {\em J. Fluid Mech.\/} {\bf 22}~(2), 385--400.

\bibitem[Sano(1981)]{Sano1980history}
{\sc Sano, T.} 1981 Unsteady flow past a sphere at low reynolds number. {\em J.
  Fluid Mech.\/} {\bf 112}, 433--441.

\bibitem[Sennitskii(1990)]{sennitskii1990self}
{\sc Sennitskii, V.~L.} 1990 Self motion of a body in a fluid. {\em J. Appl.
  Mech. Tech. Phys.\/} {\bf 31}, 266--272.

\bibitem[Shu \& Chwang(2001)]{shu2001generalized}
{\sc Shu, Jian-Jun \& Chwang, Allen~T} 2001 Generalized fundamental solutions
  for unsteady viscous flows. {\em Physical Review E\/} {\bf 63}~(5), 051201.

\bibitem[Spelman \& Lauga(2017)]{spelman2017arbitrary}
{\sc Spelman, Tamsin~A \& Lauga, Eric} 2017 Arbitrary axisymmetric steady
  streaming: Flow, force and propulsion. {\em Journal of Engineering
  Mathematics\/} {\bf 105}~(1), 31--65.

\bibitem[Visser(2011)]{Visser2011}
{\sc Visser, A.} 2011 {\em Small, wet \& rational, individual based zooplankton
  ecology\/}. {DTU Denmark}.

\bibitem[Wadhwa {\em et~al.\/}(2014)Wadhwa, Andersen \& Kiorboe]{Wadhwa2014}
{\sc Wadhwa, N., Andersen, A. \& Kiorboe, T.} 2014 Hydrodynamics and energetics
  of jumping copepod nauplii and copepodids. {\em J. Exp. Biol.\/} {\bf 217},
  3085--3094.

\bibitem[Wang \& Ardekani(2012{\natexlab{{\em a\/}}})]{wang2012inertial}
{\sc Wang, S. \& Ardekani, A.} 2012{\natexlab{{\em a\/}}} Inertial squirmer.
  {\em Phys. Fluids\/} {\bf 24}~(10), 101902.

\bibitem[Wang \& Ardekani(2012{\natexlab{{\em b\/}}})]{wang2012unsteady}
{\sc Wang, S. \& Ardekani, A.M.} 2012{\natexlab{{\em b\/}}} Unsteady swimming
  of small organisms. {\em J. Fluid Mech.\/} {\bf 702}, 286--297.

\bibitem[Wei {\em et~al.\/}(2021)Wei, Dehnavi, Aubin-Tam \&
  Tam]{wei2021measurements}
{\sc Wei, Da, Dehnavi, Parviz~G, Aubin-Tam, Marie-Eve \& Tam, Daniel} 2021
  Measurements of the unsteady flow field around beating cilia. {\em Journal of
  Fluid Mechanics\/} {\bf 915}, A70.

\bibitem[Yates(1986)]{yates1986}
{\sc Yates, G.~T.} 1986 How microorganisms move through water: The
  hydrodynamics of ciliary and flagellar propulsion reveal how microorganisms
  overcome the extreme effect of the viscosity of water. {\em Am. Sci.\/} {\bf
  74}~(4), 358--365.

\end{thebibliography}
\end{document}